\documentclass[prd,aps,twocolumn,nofootinbib,superscriptaddress,eqsecnum,floatfix,preprintnumbers,amsmath,amssymb,
nofootinbib,longbibliography]{revtex4-2}

\usepackage{amssymb,amsmath}
\usepackage{epsfig}
\usepackage[dvipsnames]{xcolor}
\usepackage[utf8]{inputenc}
\usepackage{stmaryrd}
\usepackage{mathrsfs}
\usepackage{mathalfa}
\usepackage{accents}
\usepackage[normalem]{ulem}

\usepackage{graphicx}
\graphicspath{ {c:/Documents/PhysicsNotes/GradResearch/Images/GR_Lens/} }

\newcommand{\pa}{\partial}

\newcommand{\be}{\begin{equation}}
\newcommand{\ee}{\end{equation}}
\newcommand{\bea}{\begin{eqnarray}}
\newcommand{\eea}{\end{eqnarray}}
\newcommand{\ba}{\begin{equation}\begin{aligned}}
\newcommand{\ea}{\end{aligned}\end{equation}}

\newcommand{\beg}{\begin{gather*}}
\newcommand{\eng}{\end{gather*}}
\newcommand{\hh}{,\hspace{0.5cm}}
\newcommand{\hhh}{,\hspace{0.2cm}}

\newcommand{\n}[1]{\label{#1}}
\newcommand{\bs}[1]{{\boldsymbol{#1}}}

\newcommand{\MC}[1]{{\mathcal{#1}}}

\newcommand{\CAL}{\mathcal}

\newcommand{\ts}[1]{{\boldsymbol{#1}}}
\newcommand{\ie}{\emph{i.e.} }

\def\XXint#1#2#3{{\setbox0=\hbox{$#1{#2#3}{\int}$ }
\vcenter{\hbox{$#2#3$ }}\kern-.6\wd0}}

\usepackage[makeroom]{cancel}
\usepackage[caption=false]{subfig}

\usepackage[colorlinks=true,
            citecolor=green,
            linkcolor=red,
            filecolor=cyan,
            urlcolor=magenta,
            backref=false]{hyperref}

\hypersetup{breaklinks=true}

\begin{document}

\title{Motion of a weakly charged rotating black hole in a homogeneous
electromagnetic field}

\author{Valeri P. Frolov}%
\email[]{vfrolov@ualberta.ca}
\affiliation{Theoretical Physics Institute, Department of Physics,
University of Alberta,\\
Edmonton, Alberta, T6G 2E1, Canada
}
\author{Alex Koek}
\email[]{koek@ualberta.ca}
\affiliation{Theoretical Physics Institute, Department of Physics,
University of Alberta,\\
Edmonton, Alberta, T6G 2E1, Canada
}


\begin{abstract}

In this paper we consider a rotating black hole with electric and magnetic monopole charges that moves in a static homogeneous electromagnetic field.  We assume that both the charges and the fields are weak, so that they have no effect on the spacetime geometry, which is described by the Kerr metric. We present exact solutions to Maxwell's equations describing the field of a charged rotating black hole moving in an external field background.
We use these solutions to calculate the energy, momentum, and angular momentum fluxes of the electromagnetic field into the black hole. Using these results, we obtain expressions for the torque and force acting on the moving charged rotating black hole arising as a result of its interaction with the external electromagnetic field. We calculate these quantities both in the frame comoving with the black hole and in the frame of the external background field. We use this result to derive the equations that govern the change in the black hole's mass and spin, as well as the motion of the black hole.
We provide exact solutions for specific cases which illustrate the role of charge and spin on the motion of the black hole.

 \hfill {\scriptsize Alberta Thy 6-24}
\end{abstract}

\maketitle

\section{Introduction} \label{s1}

The interaction of a black hole with an external field is an interesting theoretical problem, which in some cases allows an exact solution. An example of such a problem is the motion of a weakly charged rotating black hole in a static homogeneous electromagnetic field, which is the subject of this paper.

Let $F_{\mu\nu}$ be the strength of the static homogeneous electromagnetic field in the flat spacetime, and denote by
\be
I=F_{\mu\nu}F^{\mu\nu}\hh J=F_{\mu\nu}{}^{\star}\!F^{\mu\nu}
\ee
its two invariants. Here
\be
^{\star}\!F^{\mu\nu}=\dfrac{1}{2}e^{\mu\nu\alpha\beta}F_{\alpha\beta}
\ee
is the tensor dual to $F_{\mu\nu}$, and $e^{\mu\nu\alpha\beta}$ is the Levi-Civita tensor, $e^{0123}=-1$.
We suppose that $I\ne 0$, then there exists an inertial reference frame in which the vectors of the electric and magnetic fields are parallel. If in addition $J=0$, one of the fields vanishes in this frame.
We call this coordinate system the ``field frame" and denote it by $\tilde{K}$. Speaking about a moving black hole, we assume that such a black hole moves with some velocity $\vec{V}$ with respect to the field frame\footnote{
Let us note that if one of such $\tilde{K}$ frames is chosen, then any other frame moving with respect to it with a constant velocity directed along the common direction of the electric and magnetic fields will also be a field frame. This happens because under such motion, neither the value of the fields nor their direction changes.

}.

We assume that the electromagnetic field is weak in the following sense: We neglect the gravitational field generated by $F_{\mu\nu}$, so that the spacetime is asymptotically flat. We also assume that the black hole is weakly charged. This means that we also neglect the change of the gravitational field induced by the electromagnetic field associated with the charges. In  this setup, the gravitational field
coincides with the Kerr metric, characterized by two parameters, the mass $M$ of the black hole, and its angular momentum $J=Ma$, where $a$ is the rotation parameter.

To solve our problem, that is to find how the the electromagnetic field affects the motion of the black hole and change of its parameters, we proceed as follows. Instead of making calculations directly in the field frame $\tilde{K}$, we introduce another frame $K$ in which the black hole is at rest. In the presence of the black hole, the electromagnetic field remains homogeneous only at infinity. This asymptotic value can be obtained from the field in $\tilde{K}$ via a boost with velocity $\vec{V}$. After solving Maxwell's equations in the Kerr metric with such an asymptotic value for the electromagnetic field, one gets the desired result for the field.

This problem allows an exact solution. Let us note that the corresponding field $F_{\mu\nu}$ can be written as a sum of two fields: one is the field associated with the asymptotically homogeneous field in the presence of a neutral black hole, and the other is the field associated with the black hole's charges. An exact solution of the first problem was found by Bicak and Dvorak \cite{Bicak} (see also \cite{Bicak:85,Bicak:06,Karas:2000,Bobo} for an extended discussion of the properties of this solution).

The potential for the field of the weakly charged Kerr black hole can be written in terms of a corresponding Killing vector  \cite{Papapetrou:1966zz,Wald}. For generality, we shall consider a case where the black hole contains a magnetic monopole charge $p$ in addition to an electric charge $q$.

In 1984 Gal'tsov, Petukhov and Aliev (GPA) published a paper where they studied rotating black holes in an asymmetric electromagnetic field \cite{GALTSOV}. To describe the asymptotically homogeneous electromagnetic field in the presence of the Kerr black hole, they used an approach similar to the approach of Bicak and Dvorak \cite{Bicak}. Namely,
they used Newman-Penrose formalism and expansion of the solutions of the Teukolsky equations for the Maxwell field in terms of spin-spheroidal harmonics. To calculate the force and torque acting on the black hole, these authors used a trick similar to the one proposed by Press for the case of a black hole in a static scalar field \cite{PRESS}. Namely, they considered a thin sphere of large radius surrounding the black hole with a special distribution of the electric and magnetic currents on its surface. This distribution is chosen so that in the absence of the black hole the electromagnetic field inside the sphere is homogeneous, while outside it the field vanishes. When the black hole is put at the center of such a sphere, the field is modified and this modification results in the appearance of the force and torque acting on the shell. These quantities, when paired with a negative sign, are identified with the force and torque acting on the black hole. Using this method, GPA calculated the force acting on a charged rotating black hole in the frame where the black hole is at rest. They also derived the torque acting on a slowly rotating black hole. These results were used to describe two effects: the precession of the rotating axis of a charged black hole in an external magnetic field, and the drift of the rotating black hole induced by its interaction with an external electromagnetic field (see also \cite{UFN}).

In the present paper we consider a similar problem. Namely, we consider a weakly charged rotating Kerr black hole moving in a static homogeneous electromagnetic field. As a result of the interaction of the black hole with the external field, its velocity and spin change. Our goal is to find the equations describing the evolution of these quantities and the motion of such a black hole as it is seen by a far distant observer at rest in a specially chosen asymptotic inertial frame $\tilde{K}$. The method used for this is the following. One considers the black hole at some moment of time and introduces an asymptotically inertial frame $K$, which is instantly comoving with the black hole at this moment. One uses this Lorentz boost transformation to find the asymptotic electric and magnetic fields in the $K$ frame. Calculations of the force and torque in the $K$ frame are similar to the ones made in \cite{GALTSOV}, however there are two important differences in the methods adopted in this paper and the GPA approach
\begin{itemize}
\item We do not use the Newman-Penrose and Teukolsky equations for the Maxwell field. Instead of this with we work directly with components of the field and its stress-energy tensor in Boyer-Linduist coordinates. This simplifies the calculations and makes them more transparent.
\item We introduce vector fields generating asymptotic translations and rotations, and use them to find the asymptotic energy, momentum and angular momentum fluxes. The latter allows one to find the change of energy, momentum and spin of the black hole per unit time as seen by a distant observer, which can be identified with 4D force and torque acting on the black hole. We demonstrated that this rather simple method for the case considered in \cite{GALTSOV} gives the same results as Press' method of using fictitious shell with currents.
\end{itemize}

Using the expression for the 4D vector potential $A_{\mu}$ of the electromagnetic field in the ${K}$ frame associated with the black hole with given asymptotic values of the electric and magnetic fields, we calculated
the field strength $F_{\mu\nu}$ and the stress-energy tensor $T_{\mu\nu}$.
The latter describes in detail the distribution of the energy density and momentum fluxes, not only far away from the black hole but also in its close vicinity. The explicit expression for $T_{\mu\nu}$ is obtained  by using the GRTensor package in Maple. This expression is rather long and we do not reproduce it in the paper. It is sufficient to say that the complexity of $T_{\mu\nu}$ reflects the fact that a black hole is an extended object with a very strong gravitational field, while the black hole's rotation induces a strong frame-dragging effect in the vicinity of and within the black hole's ergoregion.

In this paper, we restrict ourselves by calculating the fluxes of the momentum and angular momentum of the electromagnetic field into the black hole. For calculations of the force and torque acting on the black hole, it is sufficient to find these fluxes at a far distance from the black hole horizon. This allows us to find the force and torque acting on the black hole in the instantly comoving frame $K$. We briefly reproduce these results in Secs.\;\ref{s2}, \;\ref{s3} and the first part of Sec.\;\ref{s4} and compare them with the results of paper \cite{GALTSOV}.

After finding the force acting on the black hole in its rest frame $K$, we use a Lorentz boost to get an expression for this force in the field frame $\tilde{K}$. At this point we make one more assumption. Namely, we assume that the change of black hole parameters $M$ and $J$, as well as the change of its velocity, is slow. In such an adiabatic approximation, the metric associated with a moving black hole momentarily coincides with the Kerr metric, in which $M$ and $J$ are the corresponding values of its slowly changing parameters.

The main new results obtained in the present paper are the folllowing
\begin{itemize}
\item It is demonstrated that in the general case, the precession of the black hole spin discussed in the paper \cite{GALTSOV} is accompanied by the decrease of its value. A general expression for the rate of this decrease is obtained.
\item  General equations describing the motion of a charged rotating black hole in an external electromagnetic field are derived, and their explicit analytic solutions are obtained for cases of special interest.
\item In particular, it is shown that the black hole spin affects the synchrotron frequency for the charged rotating black hole moving in the homogeneous magnetic field, and that this effect depends on the spin orientation.
\end{itemize}

This paper is an extension of \cite{Frolov:2024xyo}. Its main fifference is that the moving rotating black hole now has electric and magnetic monopole charges. When they vanish, our results confirm and reproduce the results of \cite{Frolov:2024xyo}. In the present paper, we mainly focus on the additional new features of the problem connected to the presence of charges.

Let us emphasise that under the adopted assumptions, the obtained equations for the evolution of mass and spin of the black hole, as well its equation of its motion, are exact. We also present their exact analytic solutions for several cases, which illustrate the role of charges and spin in the black hole's motion and change of its parameters. This makes the presented approach and the results of the paper interesting from a theoretical point of view. Possible astrophysical applications of the obtained results are briefly discussed in the last section of the paper.

The paper is organized as follows.  In Sec.\;\ref{s2} we present solutions of Maxwell's equations in the Kerr spacetime which are regular at the event horizon and describe fields which are homogeneous at infinity.  We also present solutions that describe fields generated by the black hole's electric and magnetic charges.
We designate these fields as``weak" in the sense that they do not affect the spacetime metric.  In Sec.\;\ref{s3} we calculate the energy, momentum, and angular momentum fluxes from these fields into the black hole.  Using these results, in Sec.\;\ref{s4} we analyze the evolution of the black hole's mass, angular momentum, and velocity, and derive the corresponding equations of motion. In Sec.\;\ref{s5}, we provide some interesting exact analytical solutions for special configurations which illustrate effects of the charge and spin on the black hole's motion and change of its parameters. Sec.\;\ref{s6} contains a discussion of results obtained in the paper.
To illustrate the method of calculating the force acting on a black hole immersed in an electromagnetic field, we present in the Appendix \ref{App} a similar derivation of the Lorentz force acting on a point-like charge in flat spacetime.

In this paper we will use sign conventions of the book \cite{MTW} and geometric units of $c=G=1$. We also denote 4D objects, such as 4D vectors and tensors, by boldface symbols.

\section{Weakly-charged rotating black hole in a homogeneous electromagnetic field} \label{s2}

\subsection{The Kerr Metric} \label{s2a}

The Kerr metric describing a rotating black hole in an asymptotically flat spacetime written in Boyer-Lindquist coordinates has the form
\ba\n{KERR}
ds^2 =& -\Big(1-\frac{2Mr}{\Sigma}\Big)\,dt^2 + \frac{\Sigma}{\Delta}\,dr^2+\Sigma\,d\theta^2\\
&+ \big(r^2+a^2+\frac{2Ma^{2}r}{\Sigma}\sin^2\theta\big)\sin^2\theta\,d\phi^2\\
&- \frac{4Mar}{\Sigma}\sin^2\theta\,dtd\phi \,.
\ea
Here
\be
\Sigma = r^2+a^2\cos^2\theta\hh\Delta = r^2-2Mr+a^2\, ,
\ee
where $M$ is the mass of the black hole and $a$ is its rotation parameter. For this metric $\sqrt{-g}=\Sigma \sin\theta $ and
equation $\Delta=0$ determines the location of the black hole's inner $(-)$ and outer $(+)$ event horizons, $r_{\pm}=M\pm\sqrt{M^2-a^2}$.

The Kerr black hole has two Killing vectors: vector
$\boldsymbol{\xi}$ generating time translations and vector
$\boldsymbol{\zeta}$ generating rotations
\be
\xi^{\mu}\partial_{\mu} = \partial_{t}\hh\zeta^{\mu}\partial_{\mu} = \partial_{\phi}\,.
\ee

At far distance from the black hole, the Kerr metric \eqref{KERR} reduces to the flat Minkowski metric in oblate spheroidal coordinates $(t,r,\theta,\varphi)$, which are related to the flat Cartesian coordinates $(T,X,Y,Z)$ as follows
\be\n{FLAT}
\begin{split}
&T=t\hhh X=\sqrt{r^2+a^2}\sin\theta\cos\varphi\, ,\\
&Y=\sqrt{r^2+a^2}\sin\theta\sin\varphi\hhh Z=r\cos\theta\, .
\end{split}
\ee
In this asymptotic domain, the following generators of translations along time and spatial $X$, $Y$ and $Z$ directions are defined as follows
\ba\n{TRA}
\xi^{\mu}_{(T)}\partial_{\mu} =& \partial_{t}\,,\\
\xi^{\mu}_{(X)}\partial_{\mu} =& \frac{\sqrt{r^2+a^2}}{\Sigma}\cos{\varphi}\Big(r\sin{\theta}\,\partial_{r}+\cos{\theta}\,\partial_{\theta}\Big)\\
&-\frac{\sin{\varphi}}{\sqrt{r^2+a^2}\sin{\theta}}\,\partial_{\varphi}\,,\\[5pt]
\xi^{\mu}_{(Y)}\partial_{\mu} =& \frac{\sqrt{r^2+a^2}}
{\Sigma}\sin{\varphi}\Big(r\sin{\theta}\,\partial_{r}+\cos{\theta}\,\partial_{\theta}\Big)\\
&+\frac{\cos{\varphi}}{\sqrt{r^2+a^2}\sin{\theta}}\,\partial_{\varphi}\,,\\[5pt]
\xi^{\mu}_{(Z)}\partial_{\mu} =& \frac{(r^2+a^2)\cos{\theta}}{\Sigma}\,\partial_{r}-\frac{r\sin{\theta}}{\Sigma}\,\partial_{\theta}\,.
\ea
Similarly, the generators of rotations around $X$, $Y$ and $Z$ axes are
\be\n{ROT}
\begin{split}
\zeta_{(X)}^{\mu}\pa_{\mu}&=\dfrac{\sqrt{r^2+a^2}}{\Sigma}\sin\varphi
\big[ a^2\sin\theta\cos\theta\ \pa_r-r\pa_{\theta}\big]\\
 & -\dfrac{r\cos\theta\cos\varphi}{\sqrt{r^2+a^2}\sin\theta}\ \pa_{\varphi}\, ,\\
\zeta_{(Y)}^{\mu}\pa_{\mu}&=-\dfrac{\sqrt{r^2+a^2}}{\Sigma}\cos\varphi
\big[a^2 \sin\theta\cos\theta\ \pa_r- r\pa_{\theta} \big] \\
& -\dfrac{r\cos\theta \sin\varphi }{\sqrt{r^2+a^2}\sin\theta}\ \pa_{\varphi}\, ,\\
\zeta_{(Z)}^{\mu}\pa_{\mu}&=\pa_{\varphi}\, .
\end{split}
\ee

\subsection{Asymptotically homogeneous fields} \label{s2c}

Let us consider the following vector potential for the electromagnetic field
\ba \n{AH}
A_{t}&=B_{Z}a\big(Mr(1+\cos^{2}\theta)/\Sigma-1)\\
&+\frac{aM\sin\theta\cos\theta}{\Sigma}\big[B_{X}(r\cos\psi-a\sin\psi)\\
&+B_{Y}(r\sin\psi+a\cos\psi)\big]\,,\\
A_{r}&=-(r-M)\cos\theta\sin\theta(B_{X}\sin\psi-B_{Y}\cos\psi)\,,\\
A_{\theta}&=-(ar\sin^2\theta+aM\cos^2\theta)(B_{X}\cos\psi+B_{Y}\sin\psi)\\
&+\big[r^2\cos^2\theta-(Mr-a^2)\cos(2\theta)\big](B_{X}\sin\psi-B_{Y}\cos\psi)\,,\\
A_{\phi}&=B_{Z}\sin^2\theta\big((r^2+a^2)/2-a^{2}Mr(1+\cos^2\theta)/\Sigma\big)\\
&-\sin\theta\cos\theta\big[\Delta(B_{X}\cos\psi+B_{Y}\sin\psi)\\
&+\frac{M(r^2+a^2)}{\Sigma}\big(B_{X}(r\cos\psi-a\sin\psi)\\
&+B_{Y}(r\sin\psi+a\cos\psi)\big)\big]\,.
\ea
Here $\psi$ is  given by
\be
\psi = \phi + \frac{a}{2\sqrt{M^2-a^2}}\ln\Big(\frac{r-r_{+}}{r-r_{-}}\Big)\,.
\ee
One can check that: (1) this potential is a solution of the source-free Maxwell's equations in the Kerr metric, (2) it is regular at the event horizon $r=r_+$, and (3) this four-potential asymptotically corresponds to the homogeneous magnetic field $\vec{B}=(B_X,B_Y,B_Z)$.
 The field tensor for this four-potential is
\be \n{B}
B_{\mu\nu}=\partial_{\mu}A_{\nu}-\partial_{\nu}A_{\mu}\,.
\ee

Let us apply the dual transformation to the field $B_{\mu\nu}$, and denote by $E_{\mu\nu}$ the following tensor
\be \n{BtoE}
E_{\mu\nu}={}^{\star}\!B_{\mu\nu}|_{\vec{B}\to -\vec{E}}\,.
\ee
Here
\be
^{\star}\!B_{\mu\nu} = \frac{1}{2}e_{\mu\nu\alpha\beta}B^{\alpha\beta}\,.
\ee
$e_{\mu\nu\alpha\beta}=\sqrt{-g}\epsilon_{\mu\nu\alpha\beta}$ is the Levi-Civita tensor,while $\epsilon_{\mu\nu\alpha\beta}$ is the Levi-Civita symbol with the sign convention $\epsilon_{tr\theta\phi}=1$.
The notation $\{...\}_{\vec{B}\to -\vec{E}}$
means that after the evaluation of the dual tensor $^{\star}\!B_{\mu\nu}$, one should make the following substitution in it
\be
B_{X}\rightarrow -E_{X},B_{Y}\rightarrow -E_{Y},B_{Z}\rightarrow -E_{Z}\, .
\ee

It is easy to check that this is a source-free solution of Maxwell's equations, which is regular at the event horizon and which at large distances describes the asymptotic homogeneous electric field $\vec{E}=(E_X,E_Y,E_Z)$.

For the special case when the magnetic field at far distance is parallel to the spin of the black hole, the expression \eqref{AH} greatly simplifies. Putting $B_X=B_Y=0$ in \eqref{AH}, one gets $A_r=A_{\theta}=0$ and 
\ba \n{AHZ}
A_{t}&=B_{Z}a\big(Mr(1+\cos^{2}\theta)/\Sigma-1)\, ,\\
A_{\phi}&=B_{Z}\sin^2\theta\big((r^2+a^2)/2-a^{2}Mr(1+\cos^2\theta)/\Sigma\big)\,.
\ea
It is easy to check that the corresponding potential can be written in the form \cite{Wald}
\be \n{WA}
A_{\mu}=\dfrac{1}{2}B\zeta_{\mu}+Ba\xi_{\mu}\, .
\ee

\subsection{The field of electric and magnetic monopole charges} \label{s2b}

Besides the source-free solutions described above, we consider two additional solutions. Namely, we first assume that the Kerr black hole has an electric charge $q$. A test electromagnetic field generated by this charge has the following potential \cite{Wald}
\be
A^{\mu} = -\frac{q}{2M}\xi^{\mu}\,.
\ee
The corresponding electromagnetic tensor is
\be
Q_{\mu\nu} = \pa_{\mu}A_{\nu}- \pa_{\nu}A_{\mu}\,.
\ee
Taking its dual and changing $q\to p$, one obtains the field of the monopole charge
\be
P_{\mu\nu}=\,^{\star}\!Q_{\mu\nu}|_{q\rightarrow p}\hh
^{\star}\!Q_{\mu\nu}=\frac{1}{2} e_{\mu\nu\alpha\beta}Q^{\alpha\beta}\,.
\ee
Both fields, $Q_{\mu\nu}$ and $P_{\mu\nu}$, are regular at the event horizon and decrease as $\sim 1/r^2$ as they approach infinity.
We denote by $F_{\mu\nu}$ the following superposition of all of the fields introduced above
\be \n{F}
F_{\mu\nu} = E_{\mu\nu}+B_{\mu\nu}+ Q_{\mu\nu}+P_{\mu\nu}\,.
\ee

Consider a 2D surface $\sigma$ surrounding the black hole defined by the equations $x^{\mu}=x^{\mu}(y^a)$, where $y^a=(y^1,y^2)$ are coordinates on $\sigma$. The surface area element $d\sigma_{\mu\nu}$ on $\sigma$ induced by its embedding in the spacetime is
\be
d\sigma_{\mu\nu}=e_{\mu\nu\alpha\beta} \dfrac{\pa x^{\alpha}}{\pa y^1} \dfrac{\pa x^{\beta}}{\pa y^2} dy^1 dy^2\, .
\ee
Here $e_{\mu\nu\alpha\beta}$ is the Levi-Civita tensor, $e_{0123}=\sqrt{-g}$.

The fluxes of $\ts{F}$ and $^{\star}\!\bs{F}$ through $\sigma$ determine the magnetic and electric monopole charge inside the surface. For the field $\ts{F}$, one has
\ba\n{ppqq}
\frac{1}{8\pi}\int_{\sigma} &F^{\mu\nu}\,d\sigma_{\mu\nu}=q\,,\\
\frac{1}{8\pi}\int_{\sigma}\, ^{\star}\! &F^{\mu\nu}\,d\sigma_{\mu\nu}=-p\,.
\ea

\subsection{Duality transformation}

For a given tensor $F_{\mu\nu}$, we define a new tensor $\hat{F}_{\mu\nu}$ using the following dual transformation
\be \n{DUAL}
\hat{F}_{\mu\nu}=F_{\mu\nu}\cos\phi+ ^{\star}\!\!F_{\mu\nu}\sin\phi\, .
\ee
Using the property
\be
^{\star}\! ^{\star}\!F^{\mu\nu}=-F^{\mu\nu}\, ,
\ee
one gets
\be
^{\star}\!\hat{F}_{\mu\nu}=-F_{\mu\nu}\sin\phi+ ^{\star}\!\!F_{\mu\nu}\cos\phi\, .
\ee
It is easy to check that the vacuum Maxwell equations  and
the stress-energy tensor of the field $\ts{F}$
\be\n{TTT}
\begin{split}
T_{\mu\nu} &= \frac{1}{4\pi}\big(F_{\mu\alpha}F_{\nu}^{\;\,\alpha}-\frac{1}{4}g_{\mu\nu}F_{\alpha\beta}F^{\alpha\beta}\big)\\
&=\frac{1}{8\pi}\big(F_{\mu\alpha}F_{\nu}^{\;\,\alpha}+ ^{\star}\!\!F_{\mu\alpha} {}^{\star}\!F_{\nu}^{\;\,\alpha}\big)
\,.
\end{split}
\ee
are invariant under \eqref{DUAL}.

Consider a local Lorentzian frame, and let $\ts{U}$ be a unit timelike vector of an observer at rest in this frame. Then the electric $\ts{E}$  and magnetic $\ts{B}$ fields in this frame for the strength tensor $\ts{F}$ are
\be
E^{\mu}=F^{\mu\nu}U_{\nu}\hh
B^{\mu}=-^{\star}\!F^{\mu\nu}U_{\nu}\, ,
\ee
Let us choose $\phi=\pi/2$ in \eqref{DUAL} and denote by $\hat{E}^{\mu}$ and $\hat{B}^{\mu}$ the  electric and magnetic fields for $\hat{\ts{F}}$. Then one has
\be
\hat{\ts{E}}=-\ts{B}\hh \hat{\ts{B}}=\ts{E}\, .
\ee

Under the same transformation, the electric and magnetic pole charges defined by \eqref{ppqq} become
\be
\hat{q}=-p\hh \hat{p}=q\, .
\ee
In what follows we calculate fluxes of the electromagnetic field using its stress-energy tensor,  and therefore the results of the calculations should be invariant under the change
\be \n{dual}
\ts{E}\to \ts{B}\hhh \ts{B}\to -\ts{E}\hhh q\to p\hhh p\to -q\, .
\ee

\section{Momentum and angular momentum fluxes} \label{s3}

Let $\boldsymbol{\eta}$ be a four-vector, then for the stress-energy $T_{\mu\nu}$ one can define a current $\ts{K}$
\be
K_{\mu} = T_{\mu\nu}\eta^{\nu}\,.
\ee
In what follows we assume that the current is stationary, that is $\CAL{L}_{\xi}\eta^{\mu}=0$. Let us chose a 2D surface  $S_0$ of constant radius $r=r_0$, surrounding the black hole at some time $t_0$. A set of such surfaces for the interval of time $t\in (t_0,t_0+\Delta t)$ forms a 3D surface which we denote by $\Sigma$. We denote
\be
K=\int_{\Sigma} K_{\mu} d\Sigma^{\mu}\,.
\ee
Here $d\Sigma^{\mu}$ is the 3D volume element on $\Sigma$.
We denote by $\boldsymbol{n}$ a unit spacelike vector
that is orthogonal to $\Sigma$ and directed radially outwards.

For our choice of $\Sigma$ in the Kerr spacetime, one has the following equations in Boyer-Lindquist coordinates
\be \n{NSW}
n^{\mu}=\sqrt{\frac{\Delta}{\Sigma}}\delta_r^{\mu}\hhh
d\Sigma^{\mu}=-\Delta  \delta_r^{\mu} dt d\omega\hhh d\omega=\sin\theta d\theta d\phi\,.
\ee
Then for the Kerr metric in Boyer-Lindquist coordinates one has
\be \n{KKK}
K=\CAL{K}\Delta t\hh
\mathcal{K} = -\Delta\int_{S_{0}}T_{\mu\nu}\eta^{\mu}\delta^{\nu}_{r}\,d\omega\,.
\ee
The quantity $\CAL{K}$ has a meaning of the flux density of the current $K^{\mu}$ through a 2D surface $r=r_0$ at a given moment of time. We explicitly choose the second equation in \eqref{NSW} to have a negative sign to stress that we are calculating the fluxes entering the region inside $S_0$. If $\eta^{\nu}$ is a Killing vector, then the corresponding current $K^{\mu}$ is conserved, and the flux density $\CAL{K}$ does not depend on the radius $r_0$ of the surface $S_0$.
To define the momentum and angular momentum fluxes of the electromagnetic field into the black hole,
we use expression \eqref{KKK} in which we identify $\ts{\eta}$ with vectors  $\xi_{(a)}$ and $\zeta_{(a)}$ given in \eqref{TRA} and \eqref{ROT}.

We shall use these fluxes to find the torque and the force acting on the black hole as a result of its interaction with the electromagnetic field. We assume that an observer ``studying" the black hole motion is located very far from the black hole at a distance $L\gg M$. We assume that the change of the field due to the spacetime curvature is small at a distance $\ell$ from the black hole, and that $L\gg  \ell\gg M$. To calculate the fluxes, we choose a surface $S_0$ where $r=r_0\sim \ell$.  Since the contribution of the field to the fluxes outside $r_0=\ell$ is small, it can be neglected. Formally, this means that one can take the limit $r_0\to \infty$ in \eqref{KKK}. On the other hand, the fluxes into $S_0$ change the energy, momentum and angular momentum of the  interior of $S_0$. For $L\gg \ell$ the observer at infinity interprets this change as a
result of the torque and force acting on the black hole.

Using this prescription we write
\ba
\mathcal{E}=&\lim_{r_{0}\rightarrow\infty}\bigg[\Delta\int_{S}T_{\mu\nu}\delta^{\mu}_{t}\delta^{\nu}_{r}d\omega\bigg]\,,\\
\mathcal{P}_{(a)}=&-\lim_{r_{0}\rightarrow\infty}\bigg[\Delta\int_{S}T_{\mu\nu}\xi^{\mu}_{(a)}\delta^{\nu}_{r}d\omega\bigg]\,,\\
\mathcal{J}_{(a)}=&-\lim_{r_{0}\rightarrow\infty}\bigg[\Delta\int_{S}T_{\mu\nu}\zeta^{\mu}_{(a)}\delta^{\nu}_{r}d\omega\bigg]\,.
\ea

We can break these fluxes into components due to the solely the homogeneous fields, and components that arise from the interactions between the homogeneous fields and charges
\ba
&\mathcal{E}=\mathcal{E}_{H}+\mathcal{E}_{I}\,,\\
&\mathcal{P}_{(a)}=\mathcal{P}_{H(a)}+\mathcal{P}_{I(a)}\,,\\
&\mathcal{J}_{(a)}=\mathcal{J}_{H(a)}+\mathcal{J}_{I(a)}\,,
\ea
We choose the asymptotic Cartesian coordinates so that the $Z$-axis coincides with the spin direction of the black hole, $\vec{J}=(0,0,aM)$.

The homogeneous components, which have already been calculated in \cite{Frolov:2024xyo}, are
\ba
&\mathcal{E}_{H} = 0\,,\\
&\mathcal{P}_{H(X)} = \frac{2}{3}Ma(B_{X}E_{Z}-B_{Z}E_{X})\,,\\
&\mathcal{P}_{H(Y)} = \frac{2}{3}Ma(B_{Y}E_{Z}-B_{Z}E_{Y})\,,\\
&\mathcal{P}_{H(Z)} = 0\,,
\ea
and
\ba
\mathcal{J}_{H(X)}&=\frac{2}{3}M^{2}a(B_{X}B_{Z}+E_{X}E_{Z})\\
&+\frac{22}{15}Ma^2(B_{Y}B_{Z}+E_{Y}E_{Z})\,,\\
\mathcal{J}_{H(Y)}&=\frac{2}{3}M^{2}a(B_{Y}B_{Z}+E_{Y}E_{Z})\\
&-\frac{22}{15}Ma^2(B_{X}B_{Z}+E_{X}E_{Z})\,,\\
\mathcal{J}_{H(Z)}&=-\frac{2}{3}M^{2}a(B_{X}^2+B_{Y}^{2}+E_{X}^2+E_{Y}^{2})\,.
\ea
The new interaction terms yield
\ba
&\mathcal{E}_{I} = 0\,,\\
&\mathcal{P}_{I(X)} = pB_{X}+qE_{X}\,,\\
&\mathcal{P}_{I(Y)} = pB_{Y}+qE_{Y}\,,\\
&\mathcal{P}_{I(Z)} = pB_{Z}+qE_{Z}\,,
\ea
and
\ba
&\mathcal{J}_{I(X)}=-a\big(qB_{Y}-pE_{Y}\big)\,,\\
&\mathcal{J}_{I(Y)}=a\big(qB_{X}-pE_{X}\big)\,,\\
&\mathcal{J}_{I(Z)}=0\,.
\ea

\section{Equations of Motion of a Rotating Black Hole} \label{s4}

The above relations for the fluxes are given in the frame $K$ where the black hole is at rest, and for a special choice of the asymptotic Cartesian coordinates in which the spin is directed along $Z$-axis.
It is convenient to
present these relations in a coordinate-independent form using vector notations.

\subsection{4D force}

The flux of the momentum into the black hole results in a force acting on it. The components of this force in the $K$ frame are
\ba \n{FH}
\mathcal{F}^{\mu}&=\mathcal{F}^{\mu}_H+\mathcal{F}^{\mu}_I
\hh \mathcal{F}_{H}^{t}=\mathcal{F}_{I}^{t}=0
\, ,\\
\vec{\mathcal{F}}_{H}& =- \frac{2}{3}\vec{J}\times(\vec{E}\times\vec{B})\\
&=- \frac{2}{3}[ \vec{E} (\vec{J}\cdot \vec{B})-\vec{B} (\vec{J}\cdot \vec{E})]
\, ,\\
\vec{\mathcal{F}}_{I} &= q\vec{E}+p\vec{B}\,.
\ea
Since $\mathcal{F}^{t}=0$, the mass of the black hole remains unchanged.

The force $ \vec{\mathcal{F}}_{I}$ depending on the electric and magnetic monopole charges has an expected form and coincides with the expression (24) of the paper \cite{GALTSOV}. This force is invariant under the dual transformation \eqref{dual}, as it should be. 

To explain the origin of the charge-independent force $ \vec{\mathcal{F}}_{H}$, let us consider a case when the magnetic field is parallel to the spin of the black hole, while the electric field is orthogonal to their common direction. In this case one has
\be \n{FHBJE}
\vec{\mathcal{F}}_{H}=- \frac{2}{3}BJ \vec{E}\, .
\ee
This means that the force $\vec{\mathcal{F}}_{H}$ acts on the black hole in the direction opposite to the direction of the electric field $\vec{E}$. Qualitatively this result can be explained as follows. Using the expression \eqref{WA}, one can conclude that a neutral black hole rotating in the magnetic field has an electric field  induced by rotation that is similar to the field of charge $Q=-2BJ$. The action of the electric field on this "fictitious" charge results in the force which has a form similar to \eqref{FHBJE} (see Appendix \ref{App})\footnote{
Let us note that the expression (2) of the paper \cite{GALTSOV} differes from \eqref{FHBJE} by a factor -2.}.

The force acting on the rotating black hole in its rest frame contains two components. Let us estimate their relative values. For simplicity, we put the monopole charge of the black hole $p$ equal to zero,  neglect angle dependence of the forces, and omit coefficients of the order of 1. Then one can write
\be
\mathcal{F}_{H}\sim \alpha \dfrac{G^2M^2EB}{c^4}\, .
\ee
Here we restored Newton's gravitational constant $G$ and the speed of light, $c$, and denoted by $\alpha$ the dimensionless rotation parameter of the black hole, $\alpha ={Jc}/{(GM^2)}$.
Similarly,
\be
\mathcal{F}_{I}\sim \beta {\sqrt{G} M E}\, .
\ee
Here we denote $\beta={q}/{(\sqrt{G}M)}$. For an extremal rotating black hole $\alpha=1$, while for an extremal charged black hole  $\beta=1$.

The ratio $\mathcal{F}_{H}/\mathcal{F}_{I}$ is dimensionless, and it can be presented as follows. Denote by $\epsilon_B=B^2$ the energy density of the magnetic field, and by
\be
\epsilon_{BH}=\dfrac{c^8}{G^3M^2}
\ee
the effective energy density associated with a black hole. This quantity is defined as the rest energy of the black hole $Mc^2$ divided by its "volume" $\sim R_g^3$, where $R_g=GM/c^2$.  Let us denote
\be \n{ZZZ}
\MC{Z}=\sqrt{\dfrac{\epsilon_B}{\epsilon_{BH}}}\, .
\ee
Then one has
\be
\MC{F}_H/\MC{F}_I=\dfrac{\alpha}{\beta}\MC{Z}\, .
\ee
According to our assumptions, the back reaction of the electromagnetic field on the spacetime geometry is small. This means that in our approximation one has
\be
\MC{Z}\ll 1\,.
\ee

\subsection{Torque}

Similarly, the flux of the angular momentum into the black hole results in the appearance of the torque acting on it.
The components of the torque in the $K$ frame are
\ba\n{TOR}
& \vec{\mathcal{T}}=\vec{\mathcal{T}}_H+\vec{\mathcal{T}}_I\hh\mathcal{T}_{H}=\mathcal{T}_{{1}}+\mathcal{T}_{{2}}\, ,\\
& \vec{\mathcal{T}}_{{1}} = -\frac{2M}{3}\big(\vec{B}\times(\vec{J}\times\vec{B})+\vec{E}\times(\vec{J}\times\vec{E})\big)\,,\\
&\vec{\mathcal{T}}_{{2}} =-\frac{22}{15M}\big((\vec{J}\cdot\vec{B})(\vec{J}\times\vec{B})+(\vec{J}\cdot\vec{E})(\vec{J}\times\vec{E})\big)\,,\\
&\vec{\mathcal{T}}_{I} = \frac{1}{M}\vec{J}\times\big(q\vec{B}-p\vec{E}\big)\, .
\ea
Let us stress that the expression $q\vec{B}-p\vec{E}$ which enters
 $\vec{\mathcal{T}}_{I}$ is invariant under the dual transformation \eqref{dual}, as it should be\footnote{Let us note that the expression (27) of \cite{GALTSOV} for $\vec{\mathcal{T}}_{I}$ violates the duality invariance condition. For this reason we expect that this formula contains a misprint.
We also note that in the limit of a slowly rotating black hole, that is when $a/M\ll 1$ ,  the expression \eqref{TOR} for $\vec{\mathcal{T}_{H}}$ reproduces relation (36) of \cite{GALTSOV}.}
.

One can use the equation
\be \n{FTPJ}
\frac{d\vec{J}}{d\tau}=\vec{\mathcal{T}}\,,
\ee
to find how the action of the torque changes the spin of the black hole. In this equation $\tau$ is the
proper time.

\subsection{Boosting the reference frame} \label{s3a}

The action of the force on the black hole changes its velocity. We assumed that this change is slow and used a frame momentary comoving with the black hole for the calculation of the force. To describe the black hole motion in the field frame $\tilde{K}$, we proceed as follows.
Let us recall that the asymptotic homogeneous electric and magnetic fields in the $\tilde{K}$ frame are of the form
\be
\vec{E}_{0}=E_{0}\vec{n}\hh \vec{B}_{0}=B_{0}\vec{n}\, .
\ee
Using Lorentz transformations, one finds the following components of the fields in the $K$-frame moving with respect to $\tilde{K}$ with velocity $\vec{V}$
\ba \n{EBL}
&\vec{E}=\gamma(\vec{E}_{0}+\vec{V}\times\vec{B}_{0})-(\gamma-1)\frac{\vec{V}\cdot\vec{E_{0}}}{V^2}\vec{V}\,,\\
&\vec{B}=\gamma(\vec{B}_{0}-\vec{V}\times\vec{E}_{0})-(\gamma-1)\frac{\vec{V}\cdot\vec{B_{0}}}{V^2}\vec{V}\,.
\ea
Namely, these vectors $\vec{E}$ and $\vec{B}$ should be substituted into the expressions \eqref{FH}.
After this, using that $\MC{F}^t=0$, one by means of the following Lorentz transformation obtains the 4D force  $\ts{f}=(f^0,\vec{f})$ in the $\tilde{K}$ frame
\be
f^{0} = \gamma\big(\vec{V}\cdot\vec{\mathcal{F}}\big)\hh \vec{f}=\vec{\mathcal{F}}+(\gamma-1)\frac{\vec{V}\cdot\vec{\mathcal{F}}}{V^2}\vec{V}\,.
\ee
We can separate $\ts{f}$ into contributions from the homogeneous fields and contributions from the charges as $\ts{f} = \ts{f}_{H}+\ts{f}_{I}$. Doing this, and also inserting expressions \eqref{EBL} gives us
\ba \n{allf}
f_{H}^{0} = &\frac{2}{3}\gamma^2\vec{J}\cdot\Big[(\vec{V}\cdot\vec{B}_{0})\big(\vec{E}_{0}+\vec{V}\times\vec{B}_{0}\big)\\
&-(\vec{V}\cdot\vec{E}_{0})\big(\vec{B}_{0}-\vec{V}\times\vec{E}_{0}\big)\Big]\,,\\
f_{I}^{0} = &\gamma\vec{V}\cdot\big(p\vec{B}_{0}+q\vec{E}_{0}\big)\,,\\
\vec{f}_{H} = &\frac{2}{3}\gamma\bigg[\big(\vec{B}_{0}-\vec{V}\times\vec{E}_{0}\big)\Big(\vec{J}\cdot\big(\gamma(\vec{E}_{0}+\vec{V}\times\vec{B}_{0})\\
&-(\gamma-1)\frac{\vec{V}\cdot\vec{E_{0}}}{V^2}\vec{V}\big)\Big)\\
&-\big(\vec{E}_{0}+\vec{V}\times\vec{B}_{0}\big)\Big(\vec{J}\cdot\big(\gamma(\vec{B}_{0}-\vec{V}\times\vec{E}_{0})\\
&-(\gamma-1)\frac{\vec{V}\cdot\vec{B_{0}}}{V^2}\vec{V}\big)\Big)\bigg]\,,\\
\vec{f}_{I} = &\gamma\big(p\vec{B}_{0}+q\vec{E}_{0}-\vec{V}\times(p\vec{E}_{0}-q\vec{B}_{0})\big)\,.
\ea

The 4D momentum of the black hole in the $\tilde{K}$ frame is
\be
P^{\mu}=(\gamma M, \gamma M \vec{V})\, ,
\ee
and the equations of motion of the black hole in the field frame $\tilde{K}$ are
\be
\dfrac{d(\gamma M)}{d\tau}=f^{0}\hh \dfrac{d(\gamma M\vec{V})}{d\tau}=\vec{f}\,.
\ee
Here $\tau$ is the proper time parameter. Let us recall that the black hole mass $M$ remains constant.

When the black hole does not rotate, the expressions for the torque and force acting on the black hole greatly simplify and take the form
\be
\begin{split}
&\vec{\MC{T}}=0\hh f^0=0\, ,\\
&\vec{f}=\gamma\Big[ q\vec{E}_0+p\vec{B}_0+\vec{V}\times(q\vec{B}_0-p\vec{E}_0)\Big]\, .
\end{split}
\ee
In the absence of the magnetic monopole charge, the expression for the 3D vector of the force reduces to the standard expression for the Lorentz force
\be
\vec{f}=\gamma q\Big[\vec{E}_0+\vec{V}\times\vec{B}_0 \Big]\, .
\ee
This means that a non-rotating charged black hole moves like a point-like charged particle in a static homogeneous electromagnetic field in flat spacetime.

\section{Special Cases} \label{s5}

\subsection{Spin evolution of the black hole}

Using \eqref{FTPJ} one gets
\be
\frac{d\vec{J}^2}{d\tau} = 2\vec{J}\cdot\frac{d\vec{J}}{d\tau}=2\vec{J}\cdot\vec{\mathcal{T}}\,.
\ee
Looking at \eqref{TOR}, one can see that $\vec{J}\cdot\vec{\mathcal{T}}_{H_2} = \vec{J}\cdot\vec{\mathcal{T}}_{I}=0$, and
 \be
 \vec{J}\cdot\vec{\mathcal{T}}_{H_1} = -\frac{2M}{3}\big[J^{2}B^{2}-(\vec{J}\cdot\vec{B})^2+J^{2}E^{2}-(\vec{J}\cdot\vec{E})^2\big]\,.
 \ee

Let us denote by $\vec{j}$, $\vec{e}$ and $\vec{b}$ unit vectors in the direction of $\vec{J}$, $\vec{E}$, and $\vec{B}$, respectively
\be
\vec{J}=J\vec{j}\hh \vec{E}=E\vec{e}\hh \vec{B}=B\vec{b}\, .
\ee
For a general orientation of the vectors, one has
\be
 \frac{dJ^2}{d\tau} = -\frac{4MJ^2}{3}\Big[B^2\big(1-(\vec{b}\cdot\vec{j})\big)+E^2\big(1-(\vec{e}\cdot\vec{j})\big)\Big]\,.
\ee
This relation shows that the magnitude of $J$ cannot increase. It either decreases or remains constant. The latter happens only if all three vectors $\vec{j}$,
$\vec{e}$, and $\vec{b}$ coincide. The condition $\vec{e}=\vec{b}$ means that the black hole is at rest with respect to the field frame $\tilde{K}$, or moves with respect to it with a constant velocity along the common direction of these vectors.

\subsection{Spin decrease and precession}

Let us discuss this case in more detail. We denote a common direction vector for the electric and magnetic fields by $\vec{b}$, and write
\be
\vec{B}=B\vec{b}\hh \vec{E}=E\vec{b}\hh \vec{J}=J_{\parallel}\vec{b}+\vec{J}_{\perp}\, .
\ee
Here $\vec{J}_{\perp}$ is orthogonal to the $\vec{b}$ component of the spin.
Then one has
\be
\begin{split}
& \vec{\mathcal{T}}_{{1}} = -\frac{2M}{3}(B^2+E^2)\vec{J}_{\perp}\, ,\\
&\vec{\mathcal{T}}_{{2}} =-\frac{22}{15M}
(B^2+E^2)J_{\parallel}(\vec{J}_{\perp}\times \vec{b})\, ,\\
&\vec{\mathcal{T}}_{I} = \frac{1}{M}
(qB-pE)(\vec{J}_{\perp}\times \vec{b})
\, .
\end{split}
\ee
Combining these results, one obtains the following expression for the total torque acting on the rotating black hole
\be \n{JPER}
\begin{split}
&\vec{\mathcal{T}}=\alpha \vec{J}_{\perp}+\beta (\vec{J}_{\perp}\times \vec{b})\, ,\\
&\alpha=-\frac{2M}{3}(B^2+E^2)\, ,\\
&\beta=-\frac{22}{15M}(B^2+E^2)J_{\parallel}+\frac{1}{M} (qB-pE)\, .
\end{split}
\ee
Using \eqref{FTPJ}, and the fact that $\vec{\mathcal{T}}\cdot \vec{b}=0$, one arrives at the conclusion that the  component of the black-hole's spin parallel to the field $J_{\parallel}$ is constant. This implies that the coefficients $\alpha$ and $\beta$ in \eqref{JPER} are constant. For this case, equation \eqref{FTPJ} can be easily solved exactly.

Let us choose coordinates so that the $Z$-axis coincides with the direction of the magnetic field. In these coordinates, the perpendicular component of the angular momentum vector lies in the 2D $(X,Y)$-plane
\be
\vec{J}_{\perp}=(J_X,J_Y,0)\, .
\ee
To solve the equation for the spin
\be \n{SSJJ}
\dfrac{d\vec{J}_{\perp}}{d\tau}=\alpha \vec{J}_{\perp}+\beta (\vec{J}_{\perp}\times \vec{b})\,,
\ee
we denote
\be
{\mathcal{J}}=J_X+iJ_Y\, .
\ee
Then \eqref{SSJJ} takes the form
\be
\frac{d{\mathcal{J}}}{d\tau} = \Gamma {\mathcal{J}}\hh \Gamma = \alpha-i\beta\, .
\ee
A solution of this equation is
\be\n{GGG}
{\mathcal{J}}(\tau)={\mathcal{J}}_{0}\text{exp}(\Gamma\tau)\,.
\ee
Here, ${\mathcal{J}}_0$ is the initial value of ${\mathcal{J}}$ which depends on the angular momentum of the black hole at $\tau=0$.
Equation \eqref{GGG} shows that the parameter $\alpha$ describes the rate at which the magnitude of the black hole's angular momentum decays, while the parameter $\beta$  describes the frequency at which the black-hole's spin precesses around $\vec{b}$.

\subsection{A rotating black hole moving transverse to the magnetic field} \label{s4b}

Let us consider now a case where the rotating black hole moves transverse to the magnetic field in the field frame, so that $\vec{V}\cdot\vec{B}_0=0$.
We simplify the problem by assuming that the monopole charge vanishes, $p=0$. We also assume that the electric field $\vec{E}_0$ in the field frame vanishes as well. Then relations \eqref{EBL} give
\be
\vec{B}=\gamma \vec{B}_0\hh \vec{E}=\gamma (\vec{V}\times \vec{B}_0)\, .
\ee

We consider two cases. In the first case, the spin $\vec{J}$ of the black hole is parallel to $\vec{B}_0$, and in the second case it is parallel to $\vec{V}$, and hence orthogonal to $\vec{B}_0$. To distinguish these cases we introduce a parameter $\epsilon$ which takes the value 1 for case 1, and the value 0 for case 2. Then one has
\be
\begin{split}
\vec{B}\cdot \vec{J}=\epsilon\gamma B_0J \hh\vec{E}\cdot \vec{J}=0\, .
\end{split}
\ee
Then relations \eqref{TOR} and \eqref{allf} give
\be\n{JVEQ}
\begin{split}
 & \vec{\MC{T}}=-\frac{2}{3}MB_0\gamma^2\big[ (1+ V^2) B_0 \vec{J} -\epsilon J \vec{B}_0\big]\, ,\\
 &\vec{f}=\gamma (q-\dfrac{2}{3}\epsilon\gamma B_0 J)(\vec{V}\times \vec{B}_0)\, .
\end{split}
\ee
Let us also note that one has
\be
\epsilon J \vec{B}_0=\epsilon B_0 \vec{J}\, .
\ee
This allows one to rewrite the first equation in \eqref{JVEQ} as follows
\be
\vec{\MC{T}}=-\frac{2}{3}MB_0^2\gamma^2 (1-\epsilon+ V^2) \vec{J}\,.
\ee

The second equation in \eqref{JVEQ} implies that in both cases $\vec{V}\cdot \vec{f}=0$, so that
\be
\dfrac{1}{2}M\dfrac{d(\gamma^2 V^2)}{d\tau}=
M\gamma \vec{V} \cdot \dfrac{d(\gamma \vec{V})}{d\tau}=\gamma \vec{V}\cdot \vec{f}=0\, .
\ee
Hence, both the value of the velocity $V$ and the Lorentz gamma-factor $\gamma$ are constant.

This observations allows one to solve the equation for the spin evolution
\be
\dfrac{d\vec{J}}{d\tau}=-\lambda \vec{J} \hh
\lambda=\frac{2}{3}\gamma^{2}\big(1-\epsilon+ V^2\big)B_{0}^{2}M\,.
\ee
A solution of this equation is
\be
\vec{J}=\vec{J}_{0}\,\text{exp}(-\lambda\tau)\,.
\ee
This shows that the spin of the black hole exponentially decreases and the characteristic time of this process is
\be
\tau_0=1/\lambda\, .
\ee

The equation of motion of the charged rotating black hole in the field frame is
\be
M\gamma \dfrac{d\vec{V}}{d\tau}=\vec{f}\, .
\ee

This equation is similar to \eqref{SSJJ} with $\alpha=0$. The only difference is that the parameter $\beta$ depends on time. Since $\vec{V}\cdot \vec{f}=0$, if the black hole's initial velocity $\vec{V}_0$ is orthogonal to the magnetic field $\vec{B}_0$, then it will remain orthogonal to the field for all times, so that the black hole always moves in the plane orthogonal to the field.
We chose coordinates such that the $Z$-axis coincides with the direction of the magnetic field, and we define a complex function
\be
\begin{split}
&\MC{V}=V_{X}+iV_{Y}=Ve^{-i\Phi(\tau)}\,.\\
\end{split}
\ee
Then the equation of motion
\be
M\gamma \dfrac{d\vec{V}}{d\tau}=\vec{f}
\ee
gives the following equation for $\Phi$
\be
\begin{split}
&\dfrac{d\Phi}{d\tau}=\omega\hh \omega=\frac{1}{M}\big(qB_{0}-\frac{2}{3}\epsilon\gamma B_{0}^{2}J\big)\,.
\end{split}
\ee
Solving this equation gives us
\be \n{alpha_tau}
\Phi(\tau)=-\frac{\epsilon J_0}{\gamma M^{2}V^{2}}\big(1-e^{-\lambda\tau}\big)+\frac{qB_0}{M}\tau\,.
\ee

Given a slow change in the black-hole's spin, \ie $\lambda\ll 1/M$, we see that
\be
\omega=\dfrac{d\Phi}{d\tau}\approx\frac{B_0}{M}\Big(q-\frac{2}{3}\epsilon\gamma B_{0}J_{0}\Big)\,.
\ee
This means that the black holes moves around a circular trajectory with slowly changing frequency $\omega$ (as measured by the proper time $\tau$).  Since time in the lab frame $t$ is connected with $\tau$ as $dt=\gamma d\tau$, one obtains that the angular frequency $\Omega$ as measured in the lab time is
\be \n{OM}
\Omega\approx\frac{B_0}{M\gamma}\Big(q-\frac{2}{3}\epsilon\gamma B_{0}J_{0}\Big)\,.
\ee
The second term in this relation vanishes for a non-rotating black hole, and $\Omega$ then coincides with the synchrotron frequency of a relativistic charged point-like particle with mass $M$ and charge $q$ moving in a magnetic field $B_0$ in flat spacetime.

If we assume for a moment that $\epsilon=1$, and we denote the second term in the parentheses of \eqref{OM} by $\hat{q}=\frac{2}{3}\gamma B_{0}J_{0}$, then we find that, omitting constants of order 1, the ratio $\hat{q}/q$  is proportional to the parameter $\MC{Z}$ given in \eqref{ZZZ}:
\be
\frac{\hat{q}}{q} \sim \gamma\frac{\alpha}{\beta}\mathcal{Z}\,.
\ee
This means that in the approximation adopted in this paper, the ratio $\hat{q}/{q}$ is small unless the charge of the black hole is sufficiently small. For the neutral black hole the effects connected with spin dominate.

For slow change of the spin, the angular frequency $\Omega$ also changes slowly. In this approximation, the black hole moves around a circle of radius
\be
R=\dfrac{V}{\Omega}\approx\dfrac{\gamma V M}{B_0\Big(q-\frac{2}{3}\epsilon\gamma B_{0}J_{0}\Big)}\, .
\ee
which slowly changes in time.

\subsection{Charged rotating black hole in the electric field} \label{s4a}

As a last example illustrating the motion of an electrically charged rotating black hole in a static asymptotically homogeneous field we consider a case where the magnetic field vanishes and the black hole moves in the direction of the electric field $\vec{E}_0$, For this case

\be
\begin{split}
&\vec{B}=0\hhh \vec{E}=\vec{E}_0=E_0 \vec{e}\hhh \vec{V}=V\vec{e}\, .
\end{split}
\ee
When the spin is directed along the electric field, the torque $\vec{\MC{T}}$ vanishes. When the orientation of the spin is orthogonal to $\vec{E}_0$, the torque is
\be
\vec{\MC{T}}=-\dfrac{2M}{3}E_0^2 \vec{J}\, .
\ee
This means that the spin collinear to the direction of motion is constant, while the spin orthogonal to the velocity exponentially decays. The characteristic proper time of this decay is $\tau_0=\frac{3}{2ME_0^2}$.

In both cases the 4D force $\ts{f}=(f^0,\vec{f})$ acting on the black hole is
\be
f^0=q\gamma VE_0\hh
\vec{f}=\gamma q\vec{E}_0\, .
\ee
It is easy to check that
\be
\ts{f}^2=q^2 E_0^2\,
\ee
is constant. Hence, the black hole moves along the electric field with a constant 4D acceleration $w=qE_0/M$. Let us choose the $X$ axis to be along the direction of the field and assume that initially the black hole is at rest at $X=0$. Then the equation of its motion is
\be
X=\dfrac{1}{w}[\cosh(w\tau)-1]\hh
T=\dfrac{1}{w}\sinh(w\tau)\, ,
\ee
where $\tau$ is the proper time parameter. The 3D velocity is
\be
V=\tanh(w\tau)\, .
\ee
Let us summarize. A charged black hole with charge $q$ and mass $M$ in the electric field
moves with a constant 4D acceleration, which does not depend on the black-hole's spin. This motion is the same as the motion
of a point-like particle with the same parameters. The component of the spin along the field remains the same, while the component transverse to the field decreases. The characteristic time of this decay is
\be
\tau_0=\dfrac{3}{2ME_0^2}\,.
\ee

\section{Discussion} \label{s6}

In this paper we explored the interaction between a weakly-charged rotating black hole and an external static weak homogeneous electromagnetic field.  By weak, we mean that  a back-reaction of the fields or the charges on the metric is neglected.
To calculate the force and torque acting on the moving black hole, we assumed that the change of the black-hole's parameters and its velocity is slow. We first discussed a solution of Maxwell's equations in the background of the Kerr metric which is regular at the horizon and satisfies properly chosen asymptotic conditions at the spatial infinity. Using this solution, we calculated the stress-energy tensor of the electromagnetic field arising from both the external homogeneous fields and from the electric and magnetic monopole charges. This allowed us to find the fluxes of the energy, momentum and angular momentum into the black hole. We used these results to calculate the force and torque acting on the black hole. These quantities were obtained both in the black hole frame as well in the frame of the observer at rest at asymptotic infinity.

We showed that:
\begin{itemize}
    \item For a moving charged rotating black hole in an external static homogeneous electromagnetic field, it mass remains constant.
    \item The spin of such a black hole remains constant only if (i) it is either at rest in the field frame or moves in it along the common directions of the fields, and (ii) the spin of the black hole is parallel to this common direction as well.
    \item If the above conditions are not satisfied then the black hole spin decreases.
    \item For a black hole at rest in the magnetic field, the component of its spin along the field is conserved, while its transverse component precesses and slowly decreases. The frequency of the precession depends on the black-hole's mass, charge, and spin, while the rate of the decay is proportional to the black hole's mass times the energy density of the electromagnetic field.
\end{itemize}

We also considered two other cases: (i) a black hole moving perpendicular to the magnetic field, and (ii) a black hole moving along the electric field. In the first case we demonstrated that a charged black hole moves in the plane orthogonal to the field along a circle of slowly changing radius with a slowly changing frequency, while the component of its spin orthogonal to the field slowly decreases. For a non-rotating black hole both the radius of its orbit and the frequency are constant, and coincide with the synchrotron frequency and radius of an orbiting relativistic point-like particle with the same mass and charge.

In the second case, where the charged rotating black hole moves along the electric field, we demonstrated that its 4D acceleration is constant and it does not depend on the spin orientation. The component of the spin orthogonal to the direction of motion exponentially decreases with proper time. The characteristic proper time of this process is $3/(2ME_0^2)$.

Let us emphasize that for charged rotating black holes, the effects we discussed in this paper usually contain two contributions depending on the black hole's spin and depending on its charge, respectively. Usually, the ratio of these terms is controlled by a small parameter $\MC{Z}$ (see \eqref{ZZZ}). However, when the charge $q$ is extremely small or vanishes, the spin-dependent effects dominate.
Let us note that in the case when a rotating black holes moves in a constant massless scalar field, considered in \cite{Frolov:2023gsl,Frolov:2024htj},
the equations of the motion are quite similar to the equations discussed in the present paper, with the only one important difference. The mass of the black hole moving in a scalar field can increase, and in some cases it can infinitely grow in finite time.

In a real astrophysical set-up (for example for a rotating stellar mass black hole moving around a supermassive magnetized black hole), the discussed effects connected with black hole spin are extremely small. However, if during the evolution of the early universe there were phases with strong primordial magnetic fields (see e.g. \cite{Grasso:2000wj}), these effects might be of interest.

\appendix

\section{Electric charge in a homogeneous electromagnetic field: Flat spacetime case} \label{App}

Consider a flat spacetime. Its metric in Cartesian coordinates is
\be
ds^2=\eta_{\mu\nu}dX^{\mu}dX^{\nu}=-dT^2+dX^2+dY^2+dZ^2\, .
\ee
We denote by $\ts{U}$ a four-velocity unit vector of an observer at rest in this frame. If $F_{\mu\nu}$ is a constant homogeneous electromagnetic field, then the electric field $\ts{E}$ and magnetic field $\ts{B}$ as measured by this observer are
\be
E^{\mu}=F^{\mu\nu}U_{\nu}\hh
B^{\mu}=-^{\star}\!F^{\mu\nu}U_{\nu}\, ,
\ee
where
\be
^{\star}\!F^{\mu\nu}=\dfrac{1}{2} e^{\mu\nu\alpha\beta}F_{\alpha\beta}
\ee
is the field dual to $\ts{F}$. Here $e^{\mu\nu\alpha\beta}$ is the Levi-Civita tensor, $e^{0123}=-1$.
For  a constant homogeneous electromagnetic field one has
\be\n{EB}
\begin{split}
&E^{\mu}=(0,\vec{E})\hh \vec{E}=(E_X,E_Y,E_Z)\, ,\\
&B^{\mu}=(0,\vec{B})\hh \vec{B}=(B_X,B_Y,B_Z)\, .
\end{split}
\ee
The tensor $\ts{F}$ corresponding to these vectors has the form
\be \n{FF}
F^{\mu\nu}=U^{\mu}E^{\nu}-U^{\nu}E^{\mu}-e^{\mu\nu\alpha\beta}B_{\alpha}U_{\beta}\, .
\ee

To illustrate the calculations of the force acting on a charged black hole in an external constant electromagnetic field, we present here a similar derivation of the force from an electric field acting on a charged point-like particle in a flat spacetime. We suppose that the particle is originally at rest at the origin of the frame, and denote by $q$ its charge. The electric field of such a charge is
\be
\CAL{E}^{\mu}=(0,\vec{\CAL{E}})\hh
\vec{\CAL{E}}=\dfrac{q}{r^2}\vec{n}\, .
\ee
Here $r^2=X^2+Y^2+Z^2$ and $\vec{n}=(X/r,Y/r.Z/r)$ is a unit vector directed outside a sphere of radius $r$ and orthogonal to it.
In the presence of charge, the vector $\vec{E}$ in \eqref{EB} changes to
\be
\vec{E}_q=\vec{E}+\vec{\CAL{E}}\, .
\ee
We denote by $\tilde{F}^{\mu\nu}$ the field strength \eqref{FF} obtained after the change $\vec{E}\to\vec{E}_q$ in \eqref{FF}

The stress-energy tensor of the field $\tilde{\ts{F}}$ is
\be
T_{\mu\nu}=\dfrac{1}{4\pi}\big(\tilde{F}_{\mu\alpha}\tilde{F}_{\nu}^{\ \alpha}-\dfrac{1}{4}g_{\mu\nu}\tilde{F}_{\alpha\beta}\tilde{F}^{\alpha\beta}\big)\,.
\ee
We denote by
\be
e_X^{\mu}=(0,1,0,0)\hhh e_Y^{\mu}=(0,0,1,0)\hhh e_Z^{\mu}=(0,0,0,1)
\ee
the unit vectors in the directions of the coordinate axes.
Then the components of inward momentum flux through a sphere of radius $r$ surrounding the charge are
\be \n{PP}
\vec{\CAL{P}}=r^2\int \vec{\CAL{T}} d\omega\hhh
d\omega=\sin\theta d\theta d\phi\, .
\ee
Here $d\omega$ is the surface element on a unit 2D sphere, $(\theta,\phi)$ are standard angles of the spherical coordinates, and $\vec{\CAL{T}}$ is a vector with components
$-T^{i}_{\nu}n^{\nu}$.
To find a force acting on the charge one needs only a part of this vector which depends on $q$. This part is
\be
\vec{\CAL{T}}=\dfrac{q}{4\pi r^2}\big( \vec{E}+\dfrac{q}{2r^2}\vec{n}\big)\, .
\ee
The integral of the second term in the brackets over a unit sphere vanishes, and \eqref{PP} gives
\be
\vec{\CAL{P}}=q \vec{E}\, .
\ee
By identifying $\vec{\CAL{P}}$ with the force acting on the electric charge $q$, one gets a correct expression for the force.

\section*{Acknowledgements}
This work was supported by the Natural Sciences and
Engineering Research Council of Canada. One of the authors (V.F.) is  also grateful to the Killam Trust for its financial support.


%

\end{document}